\documentclass[aps,amsfonts,unsortedaddress,nofootinbib,twocolumn]
{revtex4}
\usepackage{amsmath}
\newcommand{\lpl}{\ell_{\mbox{\tiny{pl}}}}
\begin{document}
\title{Deformed Special Relativity as an effective theory of \\measurements
on quantum gravitational backgrounds}
\author{R.  Aloisio}
\affiliation{INFN - Laboratori Nazionali
del  Gran  Sasso,  SS.  17bis,  67010  Assergi  (L'Aquila)  -  Italy}
\author{A.  Galante}
\affiliation{INFN - Laboratori Nazionali
del  Gran  Sasso,  SS.  17bis,  67010  Assergi  (L'Aquila)  -  Italy}
\affiliation{Dipartimento di Fisica,  Universit\`a di L'Aquila, Via
Vetoio  67100 Coppito (L'Aquila) - Italy}
\author{A.  Grillo}
\affiliation{INFN - Laboratori Nazionali
del  Gran  Sasso,  SS.  17bis,  67010  Assergi  (L'Aquila)  -  Italy}
\author{S. Liberati}
\affiliation{International School for Advanced Studies and INFN, Via
Beirut 2-4, 34014, Trieste - Italy}
\author{E.  Luzio}
\affiliation{Dipartimento di Fisica,  Universit\`a di L'Aquila, Via
Vetoio  67100 Coppito (L'Aquila) - Italy}
\author{F.  M\'endez}
\affiliation{INFN - Laboratori Nazionali
del Gran Sasso, SS.  17bis, 67010 Assergi (L'Aquila) - Italy}
\affiliation{Departamento de  Fisica, Universidad de  Santiago de
  Chile, Av. Ecuador 3493, Casilla 307 Stgo-2 - Chile}
%%%%%%%%%%%%%%%%%%%%%%%%%%%%%%%%%%%%%%%%%%%%%%%
\def\p{{\pi}}%
\def\ie{{\em i.e.\/}}%
\def\eg{{\em e.g.\/}}%
\def\etc{{\em etc.\/}}%
\def\etal{{\em et al.\/}}%
\def\g{{\mbox{\sl g}}}%
\def\Box{\nabla^2}%
\def\d{{\mathrm d}}%
\def\i{{\mathrm i}}%
\def\im{{\rm i}}
\def\half{{\frac{1}{2}}}
\def\L{{\mathcal L}}
\def\S{{\mathcal S}}
\def\d{{\mathrm{d}}}
\def\x{{\mathbf x}}
\def\v{{\mathbf v}}
\def\k{{\mathbf k}}
%%%%%%%%%%%%%%%%%%%%%%%%%%%%%%%%%%%%%%%%%%%%%%%
\begin{abstract}%
In this article we elaborate  on a recently proposed interpretation of
DSR   as  an  effective   measurement  theory   in  the   presence  of
non-negligible (albeit small)  quantum gravitational fluctuations.  We
provide several heuristic  arguments to explain how such  a new theory
can emerge and discuss the possible observational consequences of this
framework.
\vspace*{5mm}%
\\%
\end{abstract}%

\maketitle
%%%%%%%%%%%%%%%%%%%%%%%%%%%%%%%%%%%%%%%%%%%%%%%%%%%%%%%%%%%%%%%%%%%%%
%%%%%%%%%%%%%%%%%%%%%%%%%%%%%%%%%%%%%%%%%%%%%%%%%%%%%%%%%%%%%%%%%%%%%
\section{introduction}
%%%%%%%%%%%%%%%%%%%%%%%%%%%%%%%%%%%%%%%%%%%%%%%%%%%%%%%%%%%%%%%%%%%%%
%%%%%%%%%%%%%%%%%%%%%%%%%%%%%%%%%%%%%%%%%%%%%%%%%%%%%%%%%%%%%%%%%%%%%

Deformed or Doubly Special  Relativity (DSR) \cite{amelino1,ms} can be
understood as a  tentative to modify Special Relativity  (SR) in order
to incorporate a  new invariant scale other than  that provided by the
speed  of light $c$.  The idea  driving this  attempt is  that quantum
gravity effects seems to introduce a new dimensional fundamental scale
given by the Planck length ($\lpl$).  This is possibly problematic for
the relativity principle because  the presence of a fundamental length
scale might seem naively  incompatible with boost invariance.  However
other  ways  to  introduce  such  fundamental length  scale  could  be
envisaged  which  do  not  necessarily  lead to  a  violation  of  the
equivalence of inertial observers.

Concrete realizations of  these ideas in the momentum  space are known
\cite{ms,ame2,alu}.   In particular, deformed  boosts transformations,
deformed dispersion  relations as well  as composition laws  have been
widely   investigated   \cite{more}.    On   the   other   hand,   the
implementation of DSR in the spacetime is a more subtle subject and it
is a theme of intense debate at present time \cite{spctms}.

DSR in momentum space can be intended as a deformation of the Poincar\'e
algebra  in  the boost  sector  \cite{ms,more,luki}. Specifically  the
Lorentz commutators  among rotations and boost are  left unchanged but
the action of  boosts on momenta is changed in  a non-trivial way (see
e.g.~\cite{ms}) by corrections  which are suppressed by some
large quantum gravity scale $\kappa$. Most commonly such a scale is
taken to be  the Planck energy,
$\kappa \sim M_{\rm P} \approx 1.22\times 10^{19}\;$GeV.

It was soon recognized  ~\cite{ms,JV} that such deformed boost algebra
amounts to the assertion that  physical energy and momentum of DSR can
be   always  expressed   as  nonlinear   functions  of   a  fictitious
pseudo-momentum $\pi$,  whose components transform  linearly under the
action  of the  Lorentz group~\footnote{Indeed  this  is automatically
guaranteed if  the realization  of the Lorentz  group on  the physical
energy-momentum space  is one-to-one~\cite{LSV}.}. More  precisely one
can assume  the existence  of an invertible  map $\cal F$  between two
momentum spaces: the {\em  classical space} $\cal P$, with coordinates
$\pi_\mu$ where the Lorentz group  acts linearly and the {\em physical
space} $P$, with coordinates $p_\mu$,  where the Lorentz group acts as
the image of its action on  ${\cal P}$.  
Also, ${\cal F}$ must be such
that ${\cal  F}:[\pi_0,\vec{\pi}]\rightarrow \kappa$ for  all elements
on ${\cal P}$ with $|\vec {\pi}|=\infty$ and/or $\pi_0=\infty$.

The main  open issues in this  momentum formulation of DSR  are the so
called multiplicity  and saturation problems. The first  is related to
the fact  that in principle  there are many possible  deformations (an
infinite number, depending on the choice of an energy invariant scale,
three-momentum scale or both  \cite{luki}). This seems to suggest that
the set  of linear transformations (that  is SR) is the  only one that
have a physical  sense (it is unique and linear, what  else can we ask
for?). Moreover  the composition  law for energy  and momenta  of DSR,
being derived  by imposing a  standard composition law for  the pseudo
four-momenta $\pi_\mu$, is characterized by a saturation at the Planck
scale apparently  in open contrast with the  everyday life observation
of classical objects with transplanckian energies and momenta.

On the other hand,  since DSR is not a formulation of  QG, but gives a
set of transformations  with the typical QG scale,  it is plausible to
consider it  as a low energy limit  of QG, that is,  as some effective
theory. Indeed  such point of view  was taken in several  works on the
subject~\cite{dsrqclimit,Magueijo:2002xx,Gir,LSV}.   We hence advocate
here  the point  of view  that the  DSR transformations  (as  given in
momentum  space)  are not  fundamental  law  of transformations  among
different reference frames  but, following the proposal of~\cite{LSV},
effective relations  taking into  account the first  order corrections
due to the quantum gravitational effects.

The main purpose of this work  is to show how plausible effects due to
the quantum  nature of the space-time,  once they are  summed up, give
rise to deformed  dispersion relations of DSR-type.  In  order to show
this, in  the next sections  (II and III)  we will briefly  review the
proposal  of~\cite{LSV} formulated  in  terms of  the  metric and  the
tetrad  fields. In section  IV we  show how  does this  approach works
through a very simple  (albeit unphysical) example, but containing the
main  ideas of  the  proposal. In  the  next section  (V), we  present
heuristic arguments  for the outcome  of the average, which  will give
rise  to different  kinds of  DSR-type modifications.   Section  VI is
devoted to the analysis of the operation consequences of the framework
here discussed.  In the final  section we present the  conclusions and
also we discuss possible further developments on this topic.

%%%%%%%%%%%%%%%%%%%%%%%%%%%%%%%%%% %%%%%%%%%
\section{DSR as an effective measurement theory}
%%%%%%%%%%%%%%%%%%%%%%%%%%%%%%%%%% %%%%%%%%&&&&

Consider a four dimensional  spacetime manifold with local coordinates
$x^\mu$ and a differential line element
\begin{equation}
\label{line}
ds^2=\g_{\mu\nu}dx^\mu dx^{\nu},
\end{equation}
where $g_{\mu\nu}(x)$, is the spacetime metric.

A locally inertial frame can be defined in each point of the spacetime
through  the  set  of  four  covariant  vector  fields  $e^\alpha\,_\mu(x)$
(tetrads) defined through the relation
\begin{equation}
\label{tet}
\g_{\mu\nu}=\eta_{\alpha\beta}\,e^\beta\,_\mu e^\alpha\,_\nu.
\end{equation}
with      $\eta_{\alpha\beta}=$diag$(-1,1,1,1)$.       The     1-forms
$e^\alpha=e^\alpha\,_\mu dx^\mu$, are vectors in the cotangent space
\footnote{Here the indices
$\mu,\nu,\ldots$ are standard tensor indices, associated with a choice
of coordinates, while the indices $\alpha$, $\beta$, $\ldots$ only
label different vectors in the tetrad, and have nothing to do with any
particular chart adopted.} transforming under the action of the (local) Lorentz
group as
$
(e^\alpha) '=\Lambda^\alpha\,_\beta \,e^\beta.
$

Any vector field $V^\mu$ in the spacetime has components $V^\alpha$ in
the local inertial frame given by
$$
V^\alpha=e^\alpha\,_\mu V^\mu.
$$

Finally, the inverse tetrad, which will be denoted by $e^\mu\,_\alpha$,
is defined as the solution of
$e^\alpha\,_\mu\,e^\mu\,_\beta =\delta^\alpha\,_\beta$ and
satisfies
\begin{equation}
\label{tetinv}
\eta_{\alpha\beta}=\g_{\mu\nu}e^\mu\,_\alpha e^\nu\,_\beta.
\end{equation}
The use of a reference frame  is crucial in order to extract, from the
abstract tensors  of any  relativistic theory, scalar  quantities that
could be interpreted as measurement outcomes.

In particular, in the  usual theory of measurement \cite{defelice}, if
a  particle has  four-momentum  $\p_\mu$, its  energy  $E$ and  $i$-th
component  of three-momentum  $p_i$, measured  in the  reference frame
$\{e^\mu\,_\alpha\}$, are given by the expression
\begin{equation}
\label{lorvecp}
p_\alpha=e^\mu\,_\alpha \p_\mu,
\end{equation}
%where $\pi_0\equiv-{\cal E}$.  
Note that  now the $p_\alpha$ are  a set of four  scalars (the actual,
chart  independent,   measured  quantities  in   the  reference  frame
represented by the tetrad).

In flat  spacetime one has  $e^\mu\,_\alpha\equiv \delta^\mu\,_\alpha$
so that the  measured energy and momentum of  the particle $p_\alpha$,
will be identical to the components of $\p_\mu$, i.e.  $p_\alpha\equiv
\p_\alpha$.   In a  curved spacetime,  it is  possible to  perform the
experiment in a  locally flat space (that is,  locally inertial) or in
the accelerated frame.  In any case, quantities as $p_\alpha$ are what
we generally  associate to the outcome  of a measure.   Reduced to the
bones the proposal of~\cite{LSV} is that quantum gravitational effects
can  affect the  measurement process  and introduce  Planck suppressed
corrections to the  measured momenta with respect to  the component of
the  actual   four  momentum  of   the  particle  observed.   In  this
interpretation DSR  is an  effective theory taking  into account  in a
semiclassical  limit quantum  gravity  corrections to  the process  of
measure.  In particular one needs  to postulate that these effects act
in  such  a  way  that  the  relation  between  the  measured  scalars
$p_\alpha$ and the four momentum  of the particle $\p_\mu$ is given by
a non-linear  function of the  $\p_\mu$ and the quantum  gravity scale
$\kappa$
\begin{eqnarray}
\label{lsv}
{p}_\alpha&=&{\cal F}_\alpha[\p_\beta,\kappa],\nonumber
\\
&=&{\cal F}_\alpha[\p_\mu~{e^\mu}_\beta,\kappa].
\end{eqnarray}
For example the DSR formulation proposed by Magueijo and Smolin\,(DSR2)
\cite{ms}
\begin{eqnarray}%
&&E=\frac{-\pi_0}{1-\pi_0/\kappa}\;,%
\label{msE}\\%
&&p_i=\frac{\pi_i}{1-\pi_0/\kappa}\;,%
\label{msp}%
\end{eqnarray}%
in the measurement theory framework should actually be written as
~\cite{LSV}
\begin{equation}%
p_\alpha=\frac{\pi_\mu\,e^\mu{}_\alpha}{1 -\pi_\mu\,e^\mu{}_0/\kappa}
\;,%
\label{ms1}%
\end{equation}%
with $E=-p_0$, as usual.

Moreover a measurement of the energy-momentum of some composite object
will generally depend  also on many details of  the internal structure
of the  composite object, its  interaction with the detector,  and the
internal construction of the latter.  Let us collectively denote these
extra variables as $X$, so that%
\begin{equation}%
p_\alpha= {\cal F}_\alpha[\pi_\mu\,e^\mu{}_\beta;\kappa;X].
\label{F1a}%
\end{equation}%
In particular,  among the additional  variables $X$ one  could include
the spatial  and temporal  resolution of the  detector as well  as the
number  of particles  involved in  the measure.   In general  we shall
define an  ideal detector  a device  which will be  able to  provide a
measured quantity which is  independent of quantities intrinsic to the
detector.

It is easy to see now how, the above cited problems of DSR in momentum
space, can  be understood in this framework:  the multiplicity problem
will  be just  a  manifestation of  the  several possible  measurement
methods   which  will  in   turn  determine   different  ``measurement
functions''  ${\cal  F}_\alpha$~\cite{LSV};  the  saturation  problem,
conversely, can  be interpreted  as a constraint  only on the  size of
measurable quantum objects (given that classical objects will never be
able to  probe quantum gravitational effects)~\cite{LSV} or  it can be
even removed if  the dependence of ${\cal F}_\alpha$  on the number of
particles N  for the  composite objects is  not factorized out  but is
such that  the composite measured momenta  $p_\alpha$ saturate e.g.~at
$N\kappa$  rather than  $\kappa$. Interestingly  this one  of  the few
viable frameworks  (together with that presented  in \cite{Gir}) where
such  a dependence  on the  number of  particles of  ${\cal F}_\alpha$
(first postulated in~\cite{ms}) would be natural.

%%%%%%%%%%%%%%%%%%%%%%%%%%%
\section{Measures on a fluctuating background}
%%%%%%%%%%%%%%%%%%%%%%%%%%%

 In  order to further  explore the above  interpretative framework
for DSR one might start  from the observation that the standard theory
of  measurement heavily  relies on  the notion  of a  classical metric
structure as this is  the fundamental prerequisite for introducing the
tetrad vector  fields via  Eq.~(\ref{tet}) which in  turn characterize
the local inertial reference frames.  In a full quantum gravity theory
these tetrad fields  (or a suitable subset) are  expected to behave as
quantum fields  which only in  some appropriate limit will  define the
corresponding  classical   quantities.   
%However
When   dealing  with
measurements concerning particles at very high energies one may wonder
if the  quantum nature of  the gravitational background can  be always
safely  neglected.    In  particular  our   detectors  are  implicitly
interpreting  the  results of  their  interactions  with the  observed
particles using the classical measurement theories which is tantamount
to say that when we perform a measurement we are implicitly performing
some sort of averaging  over the quantum gravitational fluctuations in
order to recover a classical background. We shall argue here that such
unavoidable averaging  procedure will leave some energy dependent (and
Planck suppressed) corrections to the classical, low energy, metric and
hence  to the  tetrad fields.   Such  corrections could  then lead  to
Planck  suppressed  corrections  in  the measured  $p_\alpha$  scalars
defined   by~(\ref{F1a}).    Moreover    the   universal   nature   of
gravitational interactions (added to the impossibility to screen them)
leads us  to conjecture  that such corrections  will have  a universal
functional  form   (when  using  ideal  detectors)   possibly  with  a
dependence on  the kind of  particles detected and their  numbers.  In
this sense  the source  of the non  linearity in  DSR is not  some new
dynamics for  the quantum  particles of momenta  $\p_\mu$ but  a first
signal  that  our  classical  spacetime  is emergent  from  some  more
fundamental quantum theory. Let us  now try to give a more qualitative
description of our ansatz.

If  spacetime is  an intrinsic  emergent  concept, the  low energy  by
product  of  some yet  unknown  theory  of  quantum gravity,  then  at
sub-Planckian  energies it  should be  possible to  split  the quantum
operator  describing the spacetime  causal structure  ${\cal G}$  in a
classical (mean field) value
$\tilde{\g}$  plus   some  quantum  fluctuations
$\hat{h}$
\begin{equation}
\hat{\cal G}_{\mu\nu}=\tilde{\g}_{\mu\nu}+\hat{h}_{\mu\nu}
\label{split}
\end{equation}
where these  fluctuations $\hat{h}_{\mu\nu}$  can be intrinsic  to the
background metric as well as due to the presence of matter fields, but
in   any   case   characterized   by   the   quantum   gravity   scale
$\kappa$~\footnote{We stress here the strong analogy of this ansatz to
those  typical of  the so  called analog  models of  gravity  where an
emergent  geometry  is  observed  in  condensed  matter  systems  like
e.g.~Bose-Einstein condensates~\cite{becBB,  BE} (see also~\cite{ BLV}
for an extensive review on analog models).}.

In performing a measurement, as in eq.~(\ref{lorvecp}) or (\ref{lsv}),
we are  always using a classical  tetrad, and hence  a classical local
structure of  spacetime, which will  be generally the outcome  of some
averaging process over the  quantum gravitational metric.  In order to
recover Eq.~(\ref{F1a}) we have then  to postulate that the outcome of
this  average  should  depend  on  the characteristic  energy  of  the
measurement   process   as  well   as   on   $\kappa$   and  the   $X$
variables~\footnote{See  also  \cite{Magueijo:2002xx}  for  a  similar
point of  view.}.  In particular we  stress that even  if measures are
performed  in  macroscopic  detectors  the  actual  measurement  of  a
particle of  some energy $\cal E$  requires from the point  of view of
the measurement theory a  characterization of the local inertial frame
and hence of the metric on  scales at least of order $1/\cal E$. Hence
when performing  a measurement we always introduce  a classical metric
which is  supposed to  hold at the  microscopic scales over  which the
particle and our detector interact
\begin{equation}
\langle \hat{\cal G}_{\mu\nu}\rangle =\tilde{\g}_{\mu\nu}+\langle \hat{
h}_{\mu\nu}\rangle_{{\cal E},\kappa,X}
\label{med2}
\end{equation}
Compatibility with low  energy physics will be tantamount  to say that
the quantum fluctuations $\hat{h}_{\mu\nu}$  will average to zero when
the  averaging  is  done  over  scales much  larger  than  the  Planck
one. Note however that when  the averaged quantum fluctuations are non
negligible ``we  pay" the use  of a classical measurement  theory with
the  introduction   of  an  energy   dependent  ``effective  classical
metric"\footnote{See also~\cite{Smolin:2005cz} for similar ideas about
linking DSR to an energy dependent metric structure}
\begin{equation}
\langle \hat{\cal G}_{\mu\nu}\rangle =\tilde{\g}_{\mu\nu}+\langle \hat{
h}_{\mu\nu}\rangle_{{\cal E},\kappa,X}=\tilde{\g}^{\rm eff}_{\mu\nu}({
\cal E},\kappa,X)
\label{med3}
\end{equation}

Following this idea let us then rewrite Eq.~(\ref{lsv}) as an averaged
quantity
\begin{equation}
\label{med}
p_{\alpha}={\cal                 F}_{\alpha}[\p_{\beta},\kappa]=\langle
\p_{\alpha},\kappa\rangle=\p_\mu\langle e^
\mu\,_{\alpha}(x)\rangle_{{\cal E},\kappa,X}.
\end{equation}
where  the  last  line  tells  again  that  non-linearity  comes  from
averaging over  the gravitational degrees of freedom  weighted via the
energy of the particle involved ${\cal E}$, $\kappa$ and the variables
$X$. It  is interesting to  note that now  if we want to  preserve the
relativity principle we not only need to have $\kappa$ as an invariant
energy scale but also we shall  need to postulate that the form of the
quantum  gravitational  fluctuations $\hat{h}$  is  universal, in  the
sense of being the same in any inertial system of reference. 

We now  need to link the  averaged tetrad field in  (\ref{med}) to the
average of the quantum  fluctuations of the metric $\hat{h}_{\mu\nu}$.
To do so we can again split a mean value detected at very low energies
(with respect to Planck) plus  the weighted average of the fluctuating
part.
\begin{equation}
\label{medef}
\langle               e^\mu\,_{\alpha}\rangle=\tilde{e}^\mu\,_{\alpha}+
\tilde{e}^\mu\,_{\beta}\,f^{\beta} \,_{\alpha},
\end{equation}
where $f^{\beta}\,_{\alpha},$ is a matrix in the tangent spacetime and
contains     all     the     information    about     the     average;
$\tilde{e}^\mu\,_{\alpha}$ is the mean  (very low energy) value of the
tetrad    (that   in    flat   spacetime    has   the    simple   form
$\delta^\mu\,_{\alpha}$).

The inverse tetrad  $e^{\beta}\,_\mu$, is defined as the solution
  of            $           e^{\beta}\,_\mu           e^\mu\,_{\alpha}
  =\delta^{\beta}\,_{\alpha}$. However, 
since  by hypotesis,  only averaged  quantities are  available  to the
  observer,this definition  cannot be used  and the observer  can only
  compute  $\langle e^{\beta}\,_\mu\rangle$  which  satisfies $\langle
  e^{\beta}\,_\mu\rangle        \langle        e^\mu\,_{\alpha}\rangle
  =\delta^{\beta}\,_{\alpha}$. It is straightforward to show that
\begin{equation}
\label{medefinv}
\langle   e^\alpha\,_\mu   \rangle=\tilde{e}^\alpha\,_\mu-f^\alpha\,_\beta
\tilde{e}^\beta \,_\mu. 
\end{equation}

In order to match the averaged quantum fluctuations $\langle \hat{
h}_{\mu\nu}\rangle_{{\cal E},\kappa,X}$ with the part of the tetrad
dependent on $f$ we can just impose that
\begin{equation}
\langle \hat{\cal G}_{\mu\nu}\rangle =\tilde{\g}_{\mu\nu}+\langle \hat{
h}_{\mu\nu}\rangle_{{\cal    E},\kappa,X}=\eta_{\alpha\beta}   \langle
e^\alpha\,_\mu e^{\beta}\,_\mu \rangle 
\label{med3b}
\end{equation}
where we are making use of Eq.~(\ref{tet}).

As  we did  for the  evaluation of  the inverse  tetrad, here  we will
  assume  that $\langle e^\alpha\,_\mu\,e^{\beta}\,_\nu\rangle=\langle
  e^\alpha\,_\mu\rangle\langle\,e^{\beta}\,_\nu  \rangle$.
Then, the correction to the metric is
\begin{equation}
\langle                                  \hat{h}_{\mu\nu}\rangle_{{\cal
    E},\kappa,X}=-(f_{\alpha\beta}+f_{ \beta\alpha }) _{{ \cal E},\kappa,X}
\tilde{e}^\alpha\,_\mu \tilde{e}^{\beta}\,_\nu,
\label{eq:hf}
\end{equation}
Not surprisingly  in Eq.~(\ref{eq:hf}) the fluctuations  of the metric
determine only  the symmetric part  of the function $f_{\alpha\beta}$.  In what
follows, as a simplification of the  model, we shall assume  a symmetric
$f$.

So  in the end  the momentum  measured for  a particle  propagating in
spacetime  can be  read from  Eq.~(\ref{med}) using  (\ref{medef}) and
(\ref{eq:hf})
\begin{eqnarray}
p_{\alpha}&      =&      \p_\mu      \left(      \tilde{e}^\mu\,_{
\alpha}
-\frac{1}{2}\langle
\hat{h}^{\mu}\,_{\sigma}
\rangle_{{\cal E},\kappa,X}\, \tilde{e}^\sigma\,_{\alpha} \right),\nonumber
\\
&=&\p_\mu     \tilde{e}^\mu\,_{\alpha}     -     \frac{1}{2}\p^\tau
\langle \hat{h}_{\tau\sigma}\rangle_{{\cal E},\kappa,X}
\,\tilde{e}^\sigma\,_{\alpha},
\label{pgran}
\end{eqnarray}
with  $\p^\tau  =\tilde{g}^{\tau\sigma}\p_\sigma$~\footnote{Note  that
all  manipulations are  done  assuming no  corrections on  $\tilde{g}$
because  we are working  only at  first order.}.  If one  considers an
ideal detector then  one expects the variables $X$  to include at most
the number of particles involved  in the measurement process and other
quantities intrinsic of the observed  object (not of the detector). In
what follows  we shall  make the simplifying  assumption that  such an
ideal detector is used and, for the moment, we shall consider
measures of a single particle per time.

%%%%%%%%%%%%%%%%%%%%%%%%%%%%%%%%%%%%%%%%%%%%%%%%%%%%%%%%%%%%
\section{A simple analogy}
%%%%%%%%%%%%%%%%%%%%%%%%%%%%%%%%%%%%%%%%%%%%%%%%%%%%%%%%%%%%
We now try  to clarify the above proposal with  a simple, albeit quite
unphysical,  example.   Let  us  consider a  space-time  containing  a
particle and a Planck mass  Schwarzschild black hole.  We assume that,
in the frame in  which the black hole is at rest,  the particle has an
energy   much  smaller  than   Planck  energy   and  we   neglect  the
gravitational  perturbation  induced  by  the particle.   Clearly  the
particle will feel the gravitational potential of the black hole and a
global frame attached to the particle will not be inertial. However it
is  of  course always  possible  to attach  to  the  particle a  local
inertial  frame, and  the  job  is properly  done  through the  tetrad
connected to the black hole metric.

Now let us introduce an observer in the plot and assume that it is:
\begin{enumerate}
\item idealized: it does not disturb space-time in any way,
\item  coarse-grained: it cannot observe scales smaller than  some  scale
$\gg$ Planck scale,
\item not particularly clever.
\end{enumerate}

Such an observer could be tempted to define, in the frame in which the
black hole is at rest, a ``dispersion relation" for a particle of mass
$m$  using the  Schwarzschild metric  for a  Planckian black  hole. In
particular for a  particle located at a distance  $d$ much larger than
Planck length $\lpl$ from the black hole one would get
\begin{equation}  
E^2-p^2\approx m^2+4 {\frac{\lpl}{d}} E^2.
\label{eq:bhdisp}
\end{equation}
A  clever observer, even  if he/she  could not  ``see" the  black hole
(coarse-grained measures),  would easily infer  its presence (i.e.~the
fact that the spacetime is curved) by the position and time dependence
of  Eq.~(\ref{eq:bhdisp}); a not  so clever  one would  instead insist
that  spacetime is flat  and would  attach to  the particle  the above
non-trivial dispersion relation.

This situation  mimics to  some extent  that of DSR:  on one  hand the
dispersion relation is modified, on the other hand frame invariance of
physics is preserved.  In fact,  for instance it is possible (although
in general  very difficult) to find transformations  that would relate
the frame  in which the  black hole  is at rest  to that in  which the
particle  is.  An observer  at rest  with the  particle would  in fact
observe a time-varying  mass related to the time  varying potential of
the black hole at the particle location. Note also that, for this last
observer, there is  ``new", physics: e.g. if the  particle is charged,
then it will emit photons in its ``vacuum"; this ``new" physics would
be, however, frame independent.

Indeed  in  the case  of  the framework  we  are  envisaging here  the
situation is  worse than  in this example.  We don't have  a ``quantum
spacetime theory of measurement" so we are obliged to average over the
gravitational fluctuations and hence  to deal with deformed dispersion
relations similar of the form of Eq.~(\ref{eq:bhdisp}).

Of course  it is not possible  to push the just  presented analogy too
far at this  stage.  We can however complicate it a  little bit to try
to better  mimic the  real world.  For example let  us now  consider a
quantum  mechanical  description  of  the  particle so  that  it  will
naturally  have a  ``size" associated  to its  De  Broglie wave-length
$\lambda=1/E$.  If we now  assume that $\lambda$ is the characteristic
scale over  which the observer  ``measures" the particle,  then he/she
will infer a dispersion relation:
\begin{equation}
E^2-p^2 \approx m^2+4 {E^3}/{\kappa}.
\end{equation}
which is again of DSR-like form (here $\kappa=1/\lpl$).

As a final remark about the  limits of this analogy we can stress that
a single black  hole does not seem to  be an appropriate approximation
for  the QG vacuum;  to go  a step  further we  have to  consider some
ansatz for  the fluctuations of the  QG vacuum metric.  This is beyond
the scope  of the  present paper  and it will  be better  discussed in
\cite{flu2}.  However it  would seem that, in order  to preserve frame
invariance {\it \`a la} DSR,  the metric fluctuations have to preserve
some  form of  coherence in  different  frames.  In  particular it  is
probable that a necessary (although possibly not sufficient) condition
is that if  the fluctuations have a characteristic  scale, this be the
same in all frames.
%%%%%%%%%%%%%%%%%%%%%%%%%%%%%%%%%%%%%%%%%%%%%%%%%%%%%%%%%%%%%%%%%%%%%%%
\section{Heuristic arguments for the outcome of the average}
%%%%%%%%%%%%%%%%%%%%%%%%%%%%%%%%%%%%%%%%%%%%%%%%%%%%%%%%%%%%%%%%%%%%%%
In the previous section we  exposed our operative framework and linked
the  averaged fluctuations  of  the metric  to  the effective  (energy
dependent) tetrad  fields used in the standard  theory of measurement.
So doing we have seen  that the application of a classical measurement
theory on  a fluctuating background might lead  to the non-linearities
characteristic  of DSR  theories.  In  this section  we  shall further
support our hypothesis by considering some simple heuristic arguments.
%%%%%%%%%%%%%%%%%%%%%%%%%%%%%%%%%%%%%%%%%%%%%%%%%%%%%%%%%%%%%%%%%%%%%%
\subsection{Dimensional arguments}
%%%%%%%%%%%%%%%%%%%%%%%%%%%%%%%%%%%%%%%%%%%%%%%%%%%%%%%%%%%%%%%%%%%%%%
 Compatibility  with low  energy physics  implies that  the corrective
term $f$  in (\ref{medef}), must  be nearly zero when  the measurement
probes/averages  the spacetime  over distances  large  enough compared
with  the  Planck  scale,  but  becomes larger  and  larger  once  the
distances explored become smaller  and smaller. Therefore, $f$ must be
some power  of $\lpl/d$ where $d$  is the typical  distance over which
the spacetime is probed/averaged.

Let  us  assume that  the  typical scale  over  which  one probes  the
spacetime is  fixed by the wavelength  (inverse of the  energy) of the
particle detected. Then, we can  argue that $f$ should be proportional
to  $({\cal  E}/\kappa)^n$,  where   ${\cal  E}=-\p_0$  is  again  the
intrinsic energy of the particle.

The  tensor structure  of  $f$ is  what  we need  now. This  tensorial
character   can   be   constructed   with   $\eta_{\alpha\beta}$   and
$\p_{\alpha}\p_{\beta}$,  if we  insist on  covariance in  the tangent
space.  The last  option will be discarded because  all the dependence
on momenta has been decoupled by the average. Then, we will assume
\begin{equation}
f_{\alpha\beta}=\sigma \left({\cal E}/\kappa\right)^n\eta_{\alpha\beta},
\end{equation}
with $\sigma$ a dimensionless quantity  (that can include also a sign)
of the order 1. The physical momentum can be read now from (\ref{med})
and it turn out to be
\begin{equation}
\label{medf}
p_{\alpha}=\p_\mu\,e^{\mu}\,_{\alpha}\left(1+\sigma({\cal
E}/\kappa)^n\right).
\end{equation}

This  is the DSR2  relation \cite{ms}  (at first  order in  $\lpl$) if
$\alpha=1,\sigma=-1$.   Using  the  fact  that  $\eta^{\mu\nu}  \p_\mu
\p_\nu =\mu^2$ is an invariant, we find the dispersion relation 
\begin{equation}
\label{meddsr2}
E^2-{\mathbf p}^2=\mu^2\left(1+\sigma({\cal E}/\kappa)^n\right)^2.
\end{equation}

However the above is not  the most general choice, since in principle,
releasing the request  of covariance in tangent space,  we should take
into   account   all    the   possible   contributions   coming   from
$f_{00},f_{0i},f_{ij}$.   But is  not hard  to see  that contributions
from $f_{0i}$ can  be always absorbed in the other  terms, since it is
always possible  to write $|{\boldsymbol  \pi}|$ in terms of  $\cal E$
from  the  dispersion relation  $\p_\mu\p^\mu=\mu^2$.  However a  more
general tensorial structure could be given by
\begin{equation}
f_{00}=\sigma_{0}  \left({\cal E}/\kappa\right)^n,\qquad
f_{ij}=\sigma_1 \left({\cal E}/\kappa\right)^n\eta_{ij},
\end{equation}
which   in  this  case   could  accommodate,   as  the   special  case
$\sigma_{0}=-1$,   $\sigma_1=1/2$    and   $n=1$,   the    so   called
DSR1~\cite{amelino1} implementation of deformed special relativity.

In the above derivation we have guessed the result of the average over
quantum  gravitational  fluctuations  on  the base  of  a  dimensional
analysis.  However  we  shall   show  in  what  follows  that  general
assumptions about the nature of such fluctuations can also lead to the
same qualitative result.

%%%%%%%%%%%%%%%%%%%%%%%%%%%%%
\subsection{Average}
\label{sec:average}
%%%%%%%%%%%%%%%%%%%%%%%%%%%%%
Let us  review the average process with a  little more detail.  We
are   interested   in   a    definition   of   the   right   hand   of
(\ref{med2}). We can formally rewrite the average of the fluctuations
of the metric as
\begin{equation}
\langle h_{\mu\nu}\rangle_{{\cal E},\kappa} =\int dx \int{\cal D}{\cal
G}\,\hat{h}_{\mu\nu}\,{\cal P}_{{\cal E},\kappa}({\cal G}),
\end{equation}
where $\hat{h}$  is the  part of the  metric that fluctuates  (all the
effects  of  quantum gravity  are  included  here),  and the  mean  is
performed  in  the  space   of  metrics  with  weight  $\cal  P_{{\cal
E},\kappa}$. The  integration on $x$ takes into  account these effects
at some characteristic distance.

Since we  do not know how  to calculate the integral  for the metrics,
let  us  write  it  as  a  general  function  $\Phi_{\mu\nu}(x)$,  and
therefore the average has the shape
\begin{equation}
\label{eq:av}
\langle \hat{h}_{\mu\nu}\rangle=\int dx \, \Phi_{\mu\nu}(x,\kappa,{\cal
E}).
\end{equation}
Missing a  definitive theory  of quantum gravity  we shall try  to see
what can be said using a  minimal set of assumptions about the typical
fluctuations of  spacetime.  We want  to show that under  very general
assumption the  above discussed  average~(\ref{eq:av}) will lead  to a
non linear relation between measured and classical momenta.

To start with, let us make few important  assumptions about the nature
of the spacetime fluctuations.
\begin{enumerate}
\item The spacetime is characterized  by a classical  background over
which  are imposed universal quantum, Planck  scale, fluctuations.  In
order to preserve the relativity principle these fluctuations should
have the same form in any inertial system of reference.
 \item The spacetime  fluctuations have an average value  that tend to
$\pm1$ (in Planck units) for  ${\cal E}\to \kappa$ where ${\cal E}$ is
the energy  of the probing particle.  This implies that  in this limit
one cannot recover a  classical spacetime. Conversely for ${\cal E}\ll
\kappa$ we expect to recover  the classical metric so that the average
process should give vanishing fluctuations in this limit.
\item Any  measure done with particles  of energy ${\cal  E}$ probes a
large number of such  Planck scale  fluctuations. This  is a  kind of
coarse graining over a scale inversely proportional to $1/{\cal E}$.
\end{enumerate}
 
Given the above assumptions, we  can then expect that the fluctuations
are such that their average goes to zero with some power law of ${\cal
N}$ where ${\cal N}=L/\ell_{\rm  Pl}=\kappa/{\cal E}$ is the number of
fluctuations   contained   in  an   interval   of  length   $L=1/{\cal
E}$. Henceforth
\begin{equation}
\langle \hat{h}_{\mu\nu} \rangle=\int_{0}^{1/{\cal E}} {\rm d}^4 x\sqrt
{-\eta}
\,\Phi_{\mu\nu}(x,\kappa)\approx \frac{1}{{\cal N}^{4n}}=\left(
\frac{\cal E}{\kappa} \right)^{4n},
\end{equation}
and it is easy to see that in this case the modification of the dispersion
relation (\ref{pgran}) looks like
\begin{equation}
E^2-p^2=\mu^2+\frac{p^{4n+2}}{\kappa^{4n}},
\end{equation}
 which is just (\ref{meddsr2}) written in a different form.

So  for $n=1/4$  one  gets cubic  deformations. For  $n=1/2$
(Poissonian  fluctuations)  one   gets  quartic  deformations  and  so
on.  Actually it  would  be  interesting to  catalog  which kind  of
fluctuations give  the various plausible  values of $n$.  In this
way, if we shall find out that actually some form of modified relation
is  realized  in  nature, we  could  able  to  deduce which  class  of
gravitational fluctuations should be recovered in the low energy limit
of the full QG theory.  Alternatively ruling out some specific form of
dispersion relation  (within  the DSR framework)  will rule  out some
class of fluctuations of the  classical metric. Note that in principle
we could  be even  more general and  assume that the  fluctuations are
described by some function $f({\cal N})$ which goes to zero for ${\cal
N}\to \infty$.  This more general formulation should  allow to recover
all the possible modified dispersion relations associated to DSR.

%%%%%%%%%%%%%%%%%%%%%%%%%%%%%%%%%%%%%%%%%%%%%%%%%%%%%%%%%%%%%%%%%%%%%%
\subsection{Mechanical view}
%%%%%%%%%%%%%%%%%%%%%%%%%%%%%%%%%%%%%%%%%%%%%%%%%%%%%%%%%%%%%%%%%%%%%%
 An alternative  approach to the evaluation of  the outcome of the
averaging over the spacetime  quantum fluctuations is suggested by the
split of the metric (\ref{split})  which we took as a basic assumption
in  our  framework. This  splitting  is  in  fact reminiscent  of  the
linearized Einstein  equations and the analogy is  strengthened by the
fact that we are primarily considering $\tilde{g}$ as the flat metric.

In this case, however, one could argue that Einstein equations are not
fundamental in the  sense that they are not valid  once we approach to
distances of the order of the  Planck scale and moreover it could seem
inconsistent   to    apply   them    to   a   quantum    object   like
$\hat{h}_{\mu\nu}$.   In  this sense  the  analogy  with the  standard
treatment   of  dilute   Bose--Einstein   gases~\cite{becBB}  (already
stressed   when   we    introduced   (\ref{split}))   can   be   again
illuminating.  This is in  fact a  well known  example of  an analogue
model of gravity  where the background given by  the condensated atoms
behaves  at  large scales  as  an  effective  metric for  the  quantum
excitations~\cite{BE, BLV}.  Interestingly a split of a classical part
plus a quantum  fluctuation like that in (\ref{split})  is part of the
standard treatment of this systems.

In fact a  dilute Bose gas could be described  through a quantum field
${\widehat  \Psi}$  satisfying  a non-linear  Schr\"odinger  equation.
Similarly to what  we did for the metric, it is  possible to split the
quantum  field  into  a   macroscopic  (classical)  condensate  and  a
fluctuation: ${\widehat  \Psi}=\psi+{\widehat \varphi}$, with $\langle
{\widehat  \Psi}  \rangle=\psi  $.   One then  obtains  two  equations
respectively   for   the   classical   background  and   its   quantum
excitations.  The equation  for  the classical  wave  function of  the
condensate is  closed only  when the back-reaction  effect due  to the
fluctuations are  neglected.This is the  approximation contemplated by
the Gross--Pitaevskii equation.

The  interesting point  is  that when  the  back-reaction effects  are
neglected  the equations  for the  quantum perturbations  are formally
identical  to  what  one   would  have  get  from  considering  linear
perturbations   of    the   classical   background    equations   (see
e.g.~\cite{Barcelo:2003wu}).    In  strict   analogy  we   shall  here
conjecture   that  the   equations  for   the   quantum  gravitational
fluctuations are  identical to those  for the linear  perturbations of
the  classical metric,  i.e.~the linearized  Einstein  equations, when
their back-reaction is negligible.  In this analogy, therefore, it has
sense to consider  the contribution to the flat  metric of the quantum
fluctuations as a solution  of the linearized Einstein equations which
in the Lorentz gauge take the form \cite{wal}
\begin{equation}
\label{linein}
\partial_\alpha\partial^\alpha\,h _{\mu\nu}= -16\pi G_NT_{\mu\nu}\,
\end{equation}
Note also that this approach is
formally  similar to  the  ``averaged Einstein equations'' introduced  in
order   to  consider   the  inhomogeneities   at   cosmological  level
\cite{ino}, but  this is just  at formal level  and in fact,  the main
departure with this approach is the definition of the average.

At  this point we  must consider  the two  cases, namely,  presence or
absence of matter. They can be interpreted as the modifications due to
the  presence   of  matter   and  modifications  originated   just  by
fluctuations of the spacetime. Let us review first the case of vacuum.

In the radiation gauge, the formal solution of the linearized Einstein
equations is
\begin{equation}
\label{vacsol}
h_{\mu\nu}=e_{\mu\nu}e^{ikx}+e_{\mu\nu}^{*}e^{-ikx},
\end{equation}
where  $e_{\mu\nu}$ is the  polarization tensor  and $^*$  denotes the
complex  conjugate (to  render real  the previous  solution).  In this
coordinate  system (harmonic),  the  relation $2k_\mu e^\mu\,_\nu=k_\nu
e^\mu\,_\mu$ must be fulfilled as well as the condition $k^2=0$

Here we  are interested on the  result of the average  of the previous
solution, and  as we pointed out  in the introduction,  a few physical
assumptions should permit us to  give a general form for it.  Firstly,
we  assume that this  process is  ergodic so  that the  time dependent
average can  be replaced  by a  mean on ensemble.  On the  other hand,
since the temporal part has been decoupled, the only relevant piece of
the wave number is the space-like  part, which we will assume to be of
the order of  $1/\kappa$. Note that here, we are  not speaking about a
wave that propagates in a flat spacetime, instead we are talking about
the space time  itself which oscillates with spatial  amplitude of the
order of the invariant scale.

The average of the previous solution, therefore, will have the shape
\begin{equation}
\label{meanvacc}
\langle                                                      h_{\mu\nu}
\rangle=\gamma_{\mu\nu}F[x/\kappa]+\gamma_{\mu\nu}^{*}G[x/\kappa],
\end{equation}
where  $\gamma_{\mu\nu}$   is  the  result  of  the   average  of  the
polarization  tensor and  $F,G$  the  results of  the  average of  the
exponential functions of the solution.

Consider now the case  of Eq.~(\ref{linein}) with sources.  The formal
solution of the Einstein equation is
\begin{equation}
\label{lineinsol}
h_{\mu\nu}(x)=4 G_N \int_{V}
d^3y\frac{T_{\mu\nu}(y)}{
|\bf {x}-\bf{y}|},
\end{equation}
where $V$ is the past light cone of $x$.

The formal average $\langle\cdot\rangle$ of the solution is
\begin{equation}
\label{lineinsol2}
\langle h_{\mu\nu}(x)\rangle=4G_N\int
d^3y\bigg{\langle}\frac{T_{\mu\nu}(y)}{
|\bf {x}-\bf{y}|}\bigg{\rangle},
\end{equation}

Note  that the  general solution  admits  another piece  which is  the
solution of the homogeneous equation. According to our interpretation,
this means that the fluctuations of the spacetime contributes with an
additional term (\ref{meanvacc}) which  in principle could be added to
(\ref{lineinsol2}), but  which we are discarding just  to simplify the
analysis.

As  before, the  physical information  will  be put  in this  solution
through conditions on the average. A first assumption will be that the
average on time is equivalent to  space average over a large number of
copies of the system (ensembles).  Therefore the time evolution of the
system will be be replaced by the statically description averaged over
ensembles.

A  second  assumption  is  the  independence of  the  averaged  energy
momentum  tensor  from  coordinates.  That  is, we  will  assume  that
$\langle T_{\mu\nu}(x)  \rangle$ do not depend on  the coordinates. It
only depends on  the characteristic length and energy  of the particle
which is probing the space.

With this two  assumptions is clear that the  correction to the metric
has  the  same form  as  the  Newtonian  gravitational potential,  but
generated by $\langle T_{\mu\nu}\rangle$ of the particle. That is, the
correction  to the  flat  metric has  the  shape
\begin{equation}
\label{newtgrav}
\langle h_{\mu\nu} \rangle\sim G_N V \frac{\langle T_{\mu\nu}\rangle}{d},
\end{equation}
where $d$  is the  distance explored  by the particle  and $V$  is the
volume occupied by the particle. 

Note that, at  the end, all the content of QG  effects is now codified
in  the  mean value  of  $T_{\mu\nu}$  ---  which also  could  include
standard quantum  mechanics fluctuations  --- and since  these effects
are linked to  the metric fluctuations, we have to  demand that a self
consistence  between  (\ref{newtgrav})   and  the  definition  of  the
average.

In order  to get  an  explicit expression  for $\langle h_{\mu\nu}\rangle$  and
derive the  modified dispersion  relation let us  build a  (naive and
highly unrealistic, at this stage)  model of the particle as a set of
free, independent, particles of size  $l_P$ distributed on a region of
the order of $1/\cal E$, then
\begin{equation}
\label{basT}
T_{\mu\nu}=\sum_{n}{\pi_0}^{(n)}v_\mu^{(n)}v_\nu^{(n)}\delta^{
3}({x -x^{(n)}}),
\end{equation}
where $v^\mu=dx^\mu/dt$ and $x^0=t$.

It is  possible to introduce the  fluctuations as a  correction in the
velocities  due to   a   random   walk  process   (see
e.g.~\cite{Mattingly}). That is, since we are considering the particle
composed by non  interacting pieces with a mean size  of the order of
the Planck scale, we assume that  the velocity of every piece does not
depend on the other pieces and evolves randomly. Then
\begin{equation}
\label{medbasT}
\langle T_{\mu\nu}\rangle=\sum_{n}{\pi_0}^{(n)}\langle v_\mu^{(n)}
\rangle \langle v_\nu^{(n)}
\rangle \delta^{ 3}({x -x^{(n)}}).
\end{equation}

The effect  of the  average in velocities  can be written  as $\langle
v_\mu^{(n)}\rangle=\tilde{v}_\mu^{(n)}+\varpi_\mu^{(n)}$,         where
$\tilde{v}_\mu^{(n)}$ is the mean value, while $\varpi_\mu^{(n)}$, the
variance, which for  example in the random walk  process considered in
\cite{Mattingly}    turns     out    to    be    $\varpi_\mu^{(n)}\sim
\pm\sqrt{\kappa/{\cal E}^{(n)}}\delta_{0\mu}$.

Therefore, by introducing this fluctuation we find
\begin{equation}
V\langle T_{\mu\nu}\rangle={\pi_0}\left(\frac{\p_\mu}{\pi_0}+\varpi_
\mu\right) \left(\frac{\p_\nu}{\pi_0}+\varpi_\nu\right).
\end{equation}

The contribution to the metric $h$ turns out to be
\begin{equation}
\langle h_{\mu\nu}\rangle=\frac{\p_\mu \p_\nu}{\kappa^2}+
\frac{{\pi_0}\,\p_{(\mu}\,\varpi_{\nu)}}{\kappa^2},
\end{equation}
where we  have used the  approximation ${\pi_0}\,d\sim1$ and the signs $
()$ around the indices mean a symmetrized sum.

The first  term in the  RHS of the  previous equation, gives rise  to a
redefinition  of the mass,  as is  expected from  a potential  that is
purely Newtonian. The final result for $p$ is straightforward
to evaluate
\begin{equation}
\label{mattp}
p_{\alpha}=\p_{\alpha}\left(1-\frac{\mu^2}{2\kappa^2}-
\p_0\frac{\p\cdot\varpi}{2\kappa^2} \right)-
\varpi_{\alpha}\frac{\p_0\,\mu^2}{2\kappa^2}.
\end{equation}
where again $\mu^2=\eta^{\mu\nu} \p_\mu \p_\nu$ and $\p \cdot \varpi =
\p_{\alpha}\varpi^\alpha= \p_{\alpha} {\tilde{e}^{\alpha}}_\mu \varpi^
\mu$.  Note  also that energy and momenta  have different corrections,
depending,  in part,  in  the choice  of  $\varpi$ and  that our  last
expression turn out to be quadratic in momenta.

Assume  now, as  an  extra
hypothesis,  that the  linear  term in  (\ref{mattp})  gives only  the
redefinition  of mass  $\mu$,  that is  $\pi_{\alpha}\varpi^\alpha=0$.
Then   it   is   possible    to   write   $\varpi_0$   in   terms   of
$\varpi=|{\boldsymbol  \varpi}|$   and  $\pi=|{\boldsymbol  \pi}|$  as
follow  $\varpi_0=\sigma \varpi  (\pi/\pi_0)$, with  $\sigma \in[-1,1]$.
Under these assumptions we obtain for the energy and momentum
\begin{eqnarray}
\label{dsr1}
E&=&\pi_0\left(1-\frac{\mu^2}{2\kappa^2}\right)-\varpi\pi\frac{\sigma
\mu^2}{2\kappa^2},
\\
{\bf p}&=&{\boldsymbol \pi}\left(1-\frac{\mu^2}{2\kappa^2}\right)-{
\boldsymbol   \varpi}
\pi_0\frac{\mu^2}{2\kappa^2},
\end{eqnarray}
and the  first  order dispersion  relation  can  be  deduced evaluating
$E^2 -{\bf p}^2$, as was done in (\ref{meddsr2})
\begin{equation}
\label{dsrdis}
E^2-{\bf p}^2=\mu^2+\frac{\mu^2}{\kappa^2}\varpi(1-\sigma) \p_0 \p+{
\cal O}(1/
\kappa^2).
\end{equation}

A DSR1 type of dispersion relation can be obtained by setting $\varpi$
proportional to  $p$ --- equivalently, proportional to  $\pi$ since we
are working  at first order in  $\kappa^{-1}$ --- but  we also require
that this term, which is  a fluctuation, must depend on $\kappa^{-1}$.
Considering that $\varpi$ is a dimensionless quantity, we see that, in
order  to obtain  a  DSR-like  dispersion relation,  we  can make  the
(minimal) choice $\varpi\sim p/\kappa\sim\pi/\kappa$.

%%%%%%%%%%%%%%%%%%%%%%%%%%%%%%%%%%%%%%%
\section{Quantum gravity phenomenology}
%%%%%%%%%%%%%%%%%%%%%%%%%%%%%%%%%%%%%%%
We want now to discuss the operational consequences of the framework
presented here from the point of view of the quantum gravity
phenomenology tests extensively considered in the  literature \cite{
QGph}.
In fact it might seem that our proposal makes DSR a by product of a
direct measurement which would imply that  some of the processes
considered in quantum gravity phenomenology would be unaffected.
%%%%%%%%%%%%%%%%%%%%%%%%%%%%
\subsection{Time of flight}
%%%%%%%%%%%%%%%%%%%%%%%%%%%%
The time of  flight test of modified Lorentz symmetry  is based on the
cumulative effect  of the energy  dependence of the group  velocity of
photons in the presence  of dispersion relations like (\ref{meddsr2}).
By looking at the dispersion in  the time of arrival for photons which
are supposed to be emitted  simultaneously one can cast a proper bound
on the magnitude of the anomalous terms in the dispersion relation.

It might seem at first sight that such a test is lost in our framework
as  light  is  probed  only  at  its  arrival  on  Earth  and  travels
undisturbed  on long  distances.   Let us  start  by considering  that
strictly speaking  a time  of flight constraint  is obtained  with two
measurements. The  first one being the observations  of a simultaneous
emission the second  the detection of the upper bound  on the time lag
between two photons of different energy. Normally we replace the first
measurement  by a  proper assumption,  based on  our knowledge  of the
emitting object,  about the simultaneity of the  emission.  The second
measure will of course involve  the detection of photons of comparable
energies,  say  $\cal E$.   Such  measurement  hence  involves in  our
framework an extrapolation of  the local metric structure of spacetime
via an  average over  metric fluctuations on  scales of  order $1/\cal
E$. When we  do so, we can argue that we  are actually doing something
more than  simply measuring the four  momentum components $p_{\alpha}$
of the  photons arriving  on Earth.  What  we are implicitly  doing is
also  detecting   the  average  metric  that  the   photon  will  have
experienced during its  travel.  More correctly if we  assume that the
quantum fluctuations of  the metric are universal (have  the same form
at  any time  and in  any place)  then we  can effectively  ignore the
quantum interactions of gravitons with photons and simply say that the
average  effect of  the propagation  of  the photon  on a  fluctuating
background  can  be  described   as  a  photon  which  has  propagated
classically  on an energy  dependent background  as determined  by the
measurement  made on  Earth.   Of course  there  is a  caveat in  this
reasoning. We  know that  photons are red  shifted in the  travel over
cosmological distances,  hence the photons strictly  speaking will not
have probed the same effective metric  say at the start w.r.t. the end
of their trip. In principle a correct calculation would involve taking
into  account the  integrated  effect of  averaging  over the  quantum
fluctuations over different energies.
%%%%%%%%%%%%%%%%%%%%%%%
\subsection{Thresholds}
%%%%%%%%%%%%%%%%%%%%%%%
Thresholds     interactions     have     also     been     extensively
used~\cite{QGph}. In this case  the fact that the $p_{\alpha}$ scalars
are the only relevant quantities is much more clear as any interaction
can be  though as a measurement  process. Moreover it is  now obvious in
our framework  that energy and  momentum conservation will have  to be
imposed on  the classical  momenta $\pi_\mu$ as these  are actually
involved in  the interaction. However  such interaction will  now take
place on an  energy dependent background that will  lead to a modified
kinematics  for  the  observed  momenta $\p_{\alpha}$.  Studies  about
threshold   reactions   in   DSR    have   been   carried   out   (see
e.g.~\cite{Heyman:2003hs}) and the main conclusions are that
\begin{enumerate}
\item Forbidden reactions like gamma decay or vacuum threshold effect
are not allowed in DSR
\item Standard threshold reactions are very mildly modified
\end{enumerate}

We want however to comment about the impossibility in our framework to
allow in  DSR momenta space reactions  usually kinematically forbidden
in the  ``Platonic" ones. This result  is related to the  fact that if
energy moment conservation is not  satisfied in the $\pi$ variables it
will not be satisfied in  the $p$ ones~\footnote{In other words, every
solution  for  threshold  equation   in  the  classical  space  ($\pi$
variables)  can  be mapped  to  the real  space  ($p$)  and since  the
threshold  equations in  the physical  space are  maps  from threshold
equations in  the classical  space, it is  possible to see  that every
solution in the  classical space is mapped one to  one to the physical
space,  therefore,   there  are  no  new   reactions.}.  Moreover  the
possibility e.g.  of a photon decay would  be automatically associated
with a preferred system of reference.

The  interesting point  is  that  if some  energy  could be  exchanged
between  the particle  and  the gravitational  fluctuations then  this
could off set the energy balance  and indeed allow for the reaction to
happen  even  in the  $\pi$  and  consequently  it would  introduce  a
preferred system of reference!

Hence it seems that a crucial requirement for our Ansatz is that energy
exchange between gravitational and matter degrees of freedom should be
absent or negligible at the energies so far tested. Probably this is
the same to say that particles should not be so high energy that they
have a back reaction on the background metric.
%%%%%%%%%%%%%%%%%%%%%%%%%%%%%%%
\subsection{Synchrotron effect}
%%%%%%%%%%%%%%%%%%%%%%%%%%%%%%%
Synchrotron radiation has also been used~\cite{Jacobson:2002ye} to
provide constraints. The electrons responsible for such an emission
would probe a spacetime averaged on their typical energies. We then
expect them to probe similarly an effective, energy dependent,
background.

\section{Conclusions}
In this paper we have explored concrete realizations for the proposal
of~\cite{LSV}, on a new interpretation
of DSR as a new theory of the measurement. According to this new
interpretation, the {\em measured}  momenta of a particle acquires
corrections with respect to the actual four momenta of the particle,
due to  quantum gravitational effects.  Deformed dispersion relations,
as  DSR  ones,  appear  as  a  result of  this  modifications  in  the
measurements
\footnote{Note however that, given that we always work at first order,
we  cannot  reconstruct  the   explicit  form  of  $\cal{F}$  and  its
saturation properties}.

The previous idea is implemented by assuming
\begin{itemize}
\item The space time emerges as a consequence of an underlying
quantum theory of gravity.
\item Once a measurement is performed, the outcome will be  the
average on the quantum gravitational structure.
\item  The average shall depend on the characteristic energy of the
process measured, as well as on the quantum gravity scale $\kappa$ and
other possible properties (detector included) which we denote by $X$.
\item The classical structure of the spacetime is recovered when the
energies involved satisfies ${\cal E}\ll\kappa$.
\end{itemize}
With this minimal set of assumptions we have showed that for a wide
range of models for the effective structure of the underlying
spacetime, dispersion relations DSR-like can be obtained.

In all cases studied we have also assumed, as a simplicity criteria,
that {\em a)} detectors are ideal in the sense that the variable $X$
do not depend on intrinsic characteristics of apparatus and {\em b)} we
analyze  only the one-particle case.

The first model considered is  based just on dimensional analysis. The
previous requirements enforced us  to consider corrections to the flat
spacetime metric with  the shape $({\cal E}\lpl)$ and  from here it is
possible to  show that the  relation between the measured  momenta and
the particle momenta are DSR2-type. On  the base of this result is not
only the  dependence on $({\cal  E}\lpl)$, the tensorial  structure of
the   correction  ---   which   was  assumed   here  proportional   to
$\eta_{\alpha\beta}$ --- is crucial for this result.

The  natural question is  if it  is possible  to make  some statements
model  independent  about  the   result  of  the  average  on  quantum
contributions.  Indeed  this is the case  as we have  shown in section
\ref{sec:average}  under  just  a  few  assumptions.  In  fact  it  is
possible to  give a characterization  of the average which,  under our
assumptions,  is strictly related  with the  kind of  fluctuations one
considers. This result suggest that it could be possible to identify a
specific  model  of  fluctuations   with  a  specific  family  of  DSR
dispersion  relations, and then,  a dispersion  relation ruled  out by
experiments  would  force  to  discard  a specific  model  of  quantum
fluctuations  for   the  gravitational  field  (at   least  under  the
assumptions discussed in this paper).

In  the  last  model  proposed,  the  basic  assumption  is  that  the
correction to the metric can be extracted from the linearized Einstein
equations. The  main point in this  approach is that, in  spite of the
fact that Einstein  equations cannot be assumed as  fundamental at the
Planck  scale,  they  provide   a  starting  point  to  construct  the
perturbations  on the  metric.   The additional  information ---  what
makes  this approach  different  from \cite{ino}  ---  comes from  the
average which, at this point, is constructed making a few (reasonable)
physical assumptions.   In any case, this  can not be  considered as a
problem since  our goal is not  to obtain a precise  definition of the
average  from   first  principles  but   to  show  how   the  proposal
of~\cite{LSV} could work.

In fact, as we pointed out  in that section, the sense of the equation
(\ref{linein}) is not clear  unless an explicit model for fluctuations
is given through, for  example, a distribution function. However, what
are we saying  is that if we assume an ergodic  condition --- which is
an assumption on the nature of these quantum fluctuations --- and also
that  the average  gives an  effective value  for the  energy momentum
tensor, then we can approximate the solution of the equation by a
Newtonian type potential.

In  spite of  the limitations  of the  model, we  already see  how the
fluctuations  of the  spacetime  itself appear,  independently of  the
content of matter and all the dependence on the energy of the particle
that probes the space time,  comes from the definition of average. The
final result is a metric that depends on this energy.

In conclusion,  these examples show  a concrete implementation  of the
ideas of  \cite{LSV} with just a  few assumptions on  the structure of
the averages.  Note however, that in  all of them, what we have had to
do is  to argue  on the  {\em result} of  the average  and not  on the
structure itself of the spacetime. In other words, these examples show
the  compatibility of  this approach  with the  interpretation  of DSR
proposed.   Explicit models  of geometry  fluctuations leading  to DSR
will be discussed in a forthcoming paper \cite{flu2}.

 Finally,  the phenomenological issue  is addressed.  Since we  have a
link  between DSR  intended  as a  deformation  of the  outcomes of  a
measurements  due to  quantum gravitational  effects, one  could argue
that all these modifications arise only when a measurement is done and
therefore are extremely  local and completely decoupled from the
entire history of the particle. We have however argued (for example in
the  case  of the  ``time  of  flight"  observations) that  a  careful
analysis is  needed before to  draw any conclusion.  In  particular if
one  assumes  a  universal  character  for the  fluctuations  then  our
assumption that the average depend on the characteristic energy of the
process ---  that is,  the distance  scale for the  average is  of the
order of $1/{\cal E}$ ---  is everywhere valid. If, for the contrary,
the fluctuations  are not universal,  the average should  contain this
information also.  At the end this is then equivalent to replace the
metric of the space time  with an energy dependent metric, which plays
the role of the {\em effective} metric probed by the particle. In this
sense most of the constraints already discussed on DSR theories can be
recovered also in our framework.

\section*{Acknowledgements}

SL  would like  to thank  S. Sonego  for illuminating  discussions. FM
thanks INFN for postdoctoral fellowship and MECESUP, grant USA108.

%--------------------------------------------------------------

\end{document}